# Microscopic evidence of spin-driven multiferroicity and topological spin textures in monolayer NiI$_2$

Haitao Wang[1†], Tianxing Jiang[1†], Weiyi Pan[2†], Xu Wang[1], Hongyu Wang[1], Junchao Tian[1], Lianchuang Li[1,4], Dongming Zhao[1], Qingle Zhang[1], Chenxi Wang[1], Ying Yang[1], Hongjun Xiang[1,4], Changsong Xu[1,4,5]*, Donglai Feng[3,7]*, Tong Zhang[1,5,6,7]*

[1] Department of Physics, State Key Laboratory of Surface Physics and Advanced Material Laboratory, Fudan University; Shanghai, 200438, China.
[2] State Key Laboratory of Low Dimensional Quantum Physics and Department of Physics, Tsinghua University; Beijing, 100084, China.
[3] New Cornerstone Laboratory, Hefei National Laboratory; Hefei, 230088, China.
[4] Key Laboratory of Computational Physical Sciences (Ministry of Education), Institute of Computational Physical Sciences, Fudan University; Shanghai, 200433, China.
[5] Hefei National Laboratory; Hefei, 230088, China
[6] Shanghai Research Center for Quantum Sciences; Shanghai, 201315, China.
[7] Collaborative Innovation Center for Advanced Microstructures, Nanjing 210093, China

† These authors contributed to this work
* Corresponding authors. Emails: csxu@fudan.edu.cn, dlfeng@hfnl.edu.cn, tzhang18@fudan.edu.cn

**In type-II multiferroics, noncollinear spin textures are expected to induce electric polarization directly, leading to strong magnetoelectric coupling. Realizing such spin-driven multiferroicity in two-dimensional (2D) systems, and elucidating the interplay between local spins and electric polarization, are of both fundamental and technological importance. Here, using vectorial spin-polarized scanning tunneling microscopy, we investigated the spin-driven multiferroicity in monolayer NiI$_2$ at atomic scale. We identify a canted spin-spiral state with fully determined spin rotation plane, accompanied by a 2$Q$ charge modulation. At spin-spiral domain walls, we discover topological spin textures that composed of meron–antimeron pairs. These textures are associated with distinct charge pattern and notable band shifts, indicating local bound charges induced by variations of ferroelectricity at domain wall. Our observations are well captured by a realistic spin model incorporating Kitaev interactions and generalized spin-current model of type-II multiferroicity. The findings provide microscopic evidence of spin-driven multiferroicity in an extreme 2D system and establish a platform for low-dissipation, electric-field control of topological spin textures.**

The magnetoelectric coupling in multiferroic materials has opened up exciting possibilities for electric control of magnetism and vice versa [1–3]. In particular, type-II multiferroics exhibit noncollinear spin structures that break inversion symmetry, such as spin-spirals shown in Fig. 1(a). They can directly induce electric polarization and lead to strong magnetoelectric coupling [4–7]. Several mechanisms of spin-driven multiferroicity have been proposed theoretically [8–12], including spin-current model (KNB theory) [8], inverse Dzyaloshinskii-Moriya interaction [9] and the phenomenological approach based on Ginzburg-Landau theory [10]. Generally, the relation between electric polarization ($P$) and local spin textures is expressed as: $\boldsymbol{P} \propto \boldsymbol{e}_{ij} \times (\boldsymbol{S}_i \times \boldsymbol{S}_j)$, where $\boldsymbol{e}_{ij}$ is a vector pointing from $\boldsymbol{S}_i$ to $\boldsymbol{S}_j$. It highlights the critical role of spin rotation plane in generating $\boldsymbol{P}$ [Fig.1(a)]. Although type-II multiferroicity has been established through bulk magneto-electric measurements, microscopic investigations on the correlation between local spin configuration and electric polarization remain limited.

Topological spin textures, such as skyrmion and meron [Figs. 1(b-d)], represent another class of noncollinear spin structures [13,14]. They display vortex-like spin configuration and possess a nontrivial "topological charge" defined as $N = \frac{1}{4\pi} \int \boldsymbol{n} \cdot \left(\frac{\partial \boldsymbol{n}}{\partial x} \times \frac{\partial \boldsymbol{n}}{\partial y}\right) dxdy$, ($\boldsymbol{n}$ is the unit vector of spin direction). Topological spin textures can have various forms depending on different $N$, vorticity and core spin directions [Figs. 1(c,d)]. When arranged into a lattice configuration, these textures can be regarded as multi-$\boldsymbol{Q}$ spin spirals [15–17]. In metallic systems, topological spin textures can be controlled via electric current and holds great promise for spintronic applications [18,19]. Theoretically, these spin textures can also induce local electric polarization if they are generated in insulating materials, which may enable their electric-field control without joule heating. While skyrmion lattices have been observed in insulating multiferroics recently [20,21], microscopic evidence on their associated electric polarization and field controllable motion have not been reported.

NiI$_2$ is a layered van-der-Waals material [Fig. 2(a)] hosting spin-spiral and ferroelectricity in its bulk form [22,23]. Recent optical and transport studies suggest multiferroicity persists in few- and monolayer NiI$_2$ [24–27], making it promising for exploring magnetoelectric coupling at 2D limit and developing nanoscale multifunctional devices. Various theoretical works have discussed origins of multiferroicity in NiI$_2$ [28–32] and predicted the existence of topological spin textures [32,33]. However, it has been argued that optical signal may not unambiguously distinguish spin-driven multiferroicity from magnetic ordering alone [34,35], and conventional bulk magneto-electric measurements are not applicable to monolayer sample. Several STM studies have reported stripe-like modulations in monolayer NiI$_2$ [36–38], but interpretations remain divided as to whether they are spin or charge ordering. Crucially, microscopic evidence of spin-driven multiferroicity, such as direct visualization of noncollinear spin structures coupled to electric polarization remain lacking.

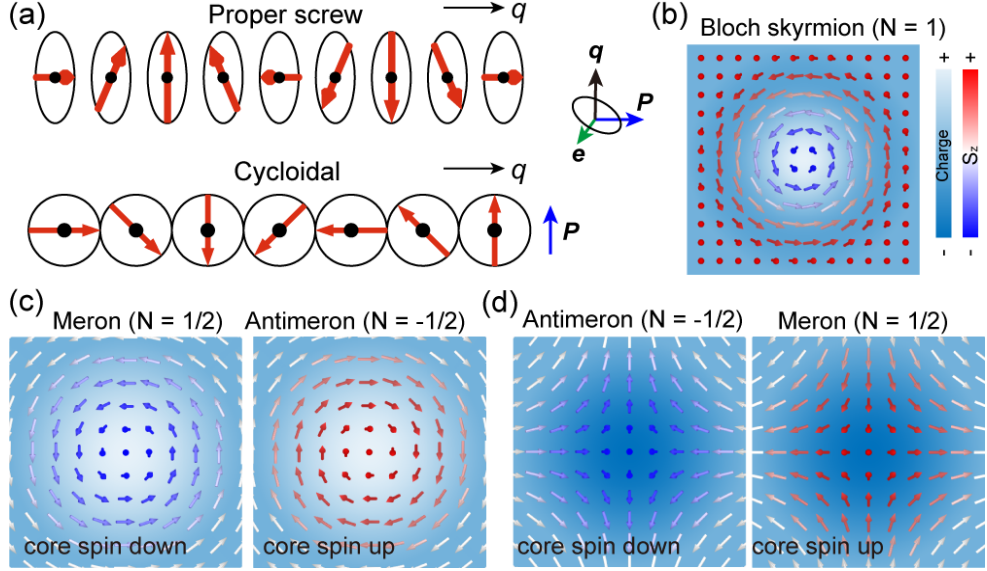

**FIG. 1.** (a) Two types of spin-spiral with different rotation plane. According to $P \propto e_{ij} \times (S_i \times S_j)$, a cycloidal type spin-spiral induces nonzero polarization $P$ but proper screw type cannot. (b) A Bloch-type skyrmion with $N=1$. (c, d) Different types of meron/antimeron with $N=\pm1/2$. Background colors in (b-d) represent bound charge density.

In this work, we employed three-axis spin-polarized scanning tunneling microscopy (SP-STM), a technique capable of resolving three-dimensional (3D) spin structures [39,40], to investigate monolayer $NiI_2$ at atomic scale. We fully resolved the 3D spin structure of the spin-spiral, characterized by a canted rotation plane and an incommensurate wavevector ($Q$). Simultaneously, we observed a $2Q$ modulation in non-spin maps, suggesting a charge modulation induced by spin-spiral. At the spin-spiral domain walls, we observed topological spin textures composed of meron/antimeron. Remarkably, each of these textures is associated with a local charge extremum and notable band shift in dI/dV spectra, indicative of local bound charges induced by the discontinuities of ferroelectric polarization. The theoretical model of type-II multiferroicity and Monte-Carlo (MC) simulation successfully reproduced these spin/charge features. Moreover, we demonstrate that the charged domain wall can be manipulated via STM tip pulses. Our study offers direct evidence of spin-driven multiferroicity in 2D limit and opens a pathway toward electric-field control of topological spin textures.

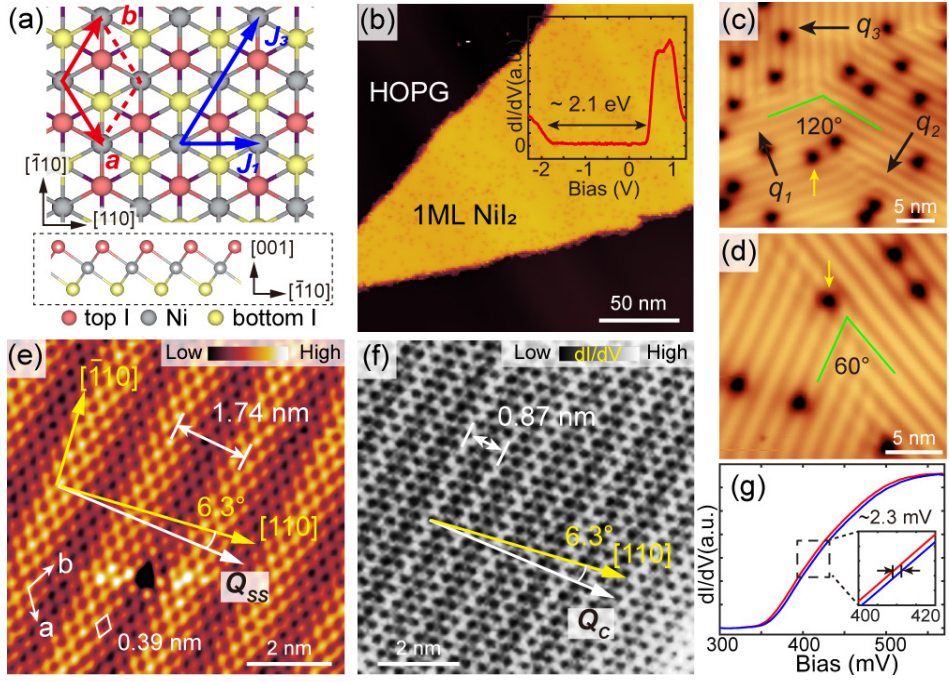

**FIG. 2.** (a) Crystal structure of monolayer NiI$_2$ (upper: top view, lower: side view). The lattice basis vectors (***a*** and ***b***), lattice directions, the first and third nearest-neighbor exchange interactions ($J_1$, $J_3$) are marked. (b) Large scale topographic image of monolayer NiI$_2$ on HOPG (V$_b$ = 1.2 V. Inset: dI/dV spectrum). (c, d) Spin-resolved images acquired by Fe-coated tip (V$_b$ =0.9 V, B$_Z$ =1 T). (e) Atomic resolved image with spin modulation (V$_b$ = 1.0 V). (f) dI/dV map taken by W tip (V$_b$ = 0.8 V), showing charge modulation. (g) dI/dV spectra taken on high/low positions of charge modulation, showing a band shift of ~ 2.3 mV (inset).

Monolayer NiI$_2$ film was epitaxially grown on the HOPG substrate [see part-I of supplementary materials, ref. 41]. Fig. 2(b) shows a large-scale STM image with atomically flat NiI$_2$ terrace. Using Fe-coated tip, we can observe stripe-like spin modulations with a period of 1.74 nm and three equivalent domains (***q***$_1$-***q***$_3$) [Figs. 2(c,d)]. They are verified as spin-spiral states as shown below. The randomly distributed "black spots" in topographic images are induced by point defects (see Fig. S1 of ref. 41).

The inset of Fig. 2(b) presents typical dI/dV spectrum taken at defect-free area, revealing an insulating gap of ~2.1 eV and a pronounced conductance peak at 0.7–1.0 eV. Our DFT calculations of local density-of-state (LDOS) reproduce these features and indicates the peak is mainly contributed by Ni orbitals with atomic-site dependent spin polarization [Fig. S13(b,c) [41]]. Fig. 2(e) displays an atomically resolved image with spin resolution acquired within this energy range. The measured lattice constant of 0.39 nm is consistent with bulk NiI$_2$ [22]; while the spin-modulation wavevector lies close to [110] direction with a small deviation of 6.3°, giving a wavevector of ***Q***$_{SS}$ = (0.13, 0.09, 0) in units of reciprocal vectors (see Fig. S2 for details [41]). This differs from bulk NiI$_2$ whose in-plane ***Q*** vector lies along [1$\bar{1}$0]. Such difference can be captured by a realistic spin model [32,42] incorporating Kitaev interaction shown later.

In addition to spin modulation, we also observed a charge modulation ($Q_C$) with half the wavelength of spin modulation ($Q_C = 2Q_{SS}$) in non-spin sensitive dI/dV map [Fig. 2(f)]. This modulation has not been reported in previous studies [36–38]. dI/dV spectra reveal a small band shift of ~2.3 mV between high/low intensity positions of this modulation [Fig. 2(g); see also Fig. S9 [41]]. We shall note that a $2Q$ modulation induced by spin spiral has been observed in magnetic metal films such as Mn/W(001) [43,44], which was attributed to spin-orientation dependent LDOS modulation induced by spin-orbital coupling [45]. Meanwhile, a phenomenological theory of type-II multiferroicity [10] predicts that a spin-spiral can generate modulated polarization with $2Q$ via $\boldsymbol{P} \propto \boldsymbol{M} \times (\nabla \times \boldsymbol{M})$. Our DFT calculations presented later suggest both mechanisms may apply.

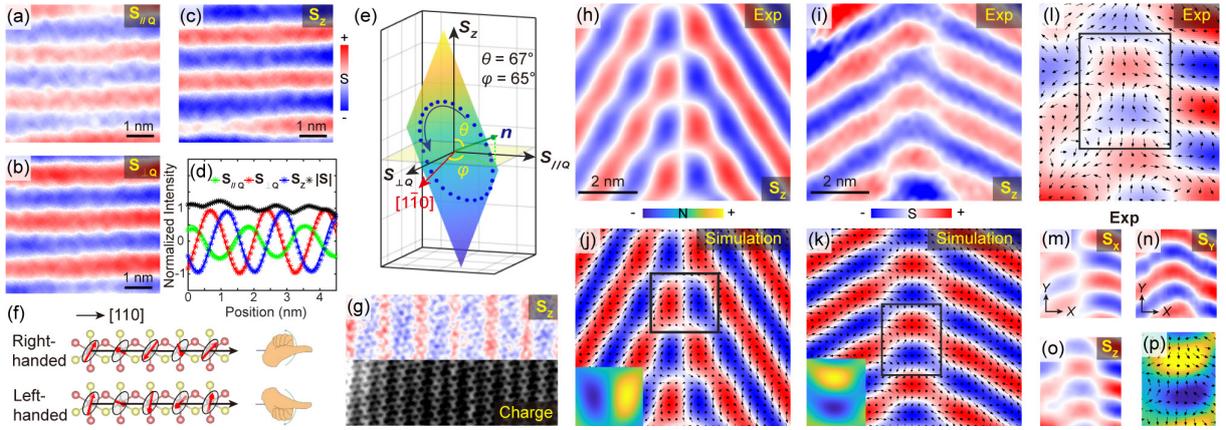

**FIG. 3.** (a-c) Spin maps with sensitivity along $\boldsymbol{Q}$, in-plane perpendicular to $\boldsymbol{Q}$ and along Z, respectively. (d) The line profiles of panels (a-c), averaged along the stripes, and the total spin magnitude |$\boldsymbol{S}$|. (e) The determined spin rotation plane in spin-space, $\boldsymbol{n}$ is the normal direction. (f) Measured real-space spin structure of right-/left- handed spin-spiral domains of 1ML NiI$_2$. (g) S$_Z$ and charge map taken at the same area. (h, i) Measured S$_Z$ map of 60°/120° domain walls, respectively. (j, k) Simulated S$_Z$ map of 60°/120° domain walls by superposition of two spin-spirals. (inset: topological charge density map). (l-o) The measured 3D spin structure of 120° domain wall and the S$_x$, S$_y$, S$_z$ maps. (p) topological charge density of the meron/antimeron pair marked in (l).

We next examine the detailed structure of spin modulations. Using an Fe-coated tip magnetized by vectorial magnetic field [40], we measured pure spin components along three orthogonal axes. Non-spin signals were eliminated by reversing the tip magnetization, and the full 3D spin structure can be constructed (see part-II-3 of ref. 41). Figs. 3(a-c) show the spin component maps along $\boldsymbol{Q}$ ($S_{\parallel Q}$), in-plane perpendicular to $\boldsymbol{Q}$ ($S_{\perp Q}$) and out-of-plane ($S_Z$) directions. Spin modulations are observed in all these maps, but with different phases and intensities [Fig. 3(d)]. The nearly constant spin amplitude, $|S| = \sqrt{S_{\parallel Q}^2 + S_{\perp Q}^2 + S_Z^2}$, suggest the modulation comes from rotation of local spins, i.e., a spin-spiral state [46]. A 3D view in spin-space [Fig. 3(e)] reveals the spin rotation plane, which is canted from $\boldsymbol{Q}_{SS}$ with polar/azimuthal angles of $\theta = 67°$ and $\varphi = 65°$. The real-space spin structure shows that this spin spiral has a right-handed chirality [Fig. 3(f) upper panel, see also Fig. S4 [41]]; while a left-handed domain with similar rotation plane was also observed [Fig. 3(f) lower panel, see also Fig. S5 [41]], suggesting a chirality degeneracy in NiI$_2$.

Meanwhile, the 2$\boldsymbol{Q}$ charge modulation can be resolved by adding up dI/dV maps with opposite tip magnetization, which eliminates spin signal (Fig. S8 [41]). As shown in Fig.3(g), the S$_Z$ and charge modulation are nearly in phase, i.e, the extrema of charge modulation align with the maxima of |S$_z$|. This relation is reproduced by our DFT calculation.

Next, we investigate domain-wall structure of spin spirals. As shown in Figs. 2(c,d), spin modulation along different orientation can intersect at angles of 60° or 120°, forming domain walls we refer to as 60°/120° walls. High-resolution S$_Z$ maps around these domain walls [Figs. 3(h,i)] reveal distinct spin textures, appearing as superposition of two spin-spirals (a multi-$\boldsymbol{Q}$ state [47–50]). To gain intuitive understanding, we first simulated the domain walls by superposing two spin spirals with exponentially decaying amplitudes across the boundary (Part-III of SM [41]). The simulated S$_Z$ components [Figs. 3(j,k)] well match the experimental data in Figs. 3(h,i). Meanwhile, the in-plane spin components at the domain wall region resemble meron/antimeron structures shown in either Fig. 1(c) or 1(d) (depends on the chirality of two spin spirals, see Fig. S11[41]), and exhibit non-zero topological charge densities (insets).

To further verify these topological spin textures, we measured all three spin components (S$_X$, S$_Y$, S$_Z$) across a 120° wall [Figs.3(m-o)]. The obtained 3D spin structure [Fig. 3(*l*)] directly shows that meron-antimeron pairs formed at the domain wall, with structures similar to Fig. 1(d) (see Fig. S7 for more details [41]). The derived topological charge density [Fig. 3(p)] also matches the simulation in Fig. 3(k).

Having determined the detailed spin structure, we turn to examine the multiferroicity in monolayer NiI$_2$. If a spin spiral can induce ferroelectricity, the domain walls will break the continuity of electric polarization $\boldsymbol{P}$ and induce localized bound charges. Figs. 4(a,b) present two non-spin-sensitive dI/dV maps taken around 60°/120° walls. Remarkably, distinct charge patterns with much stronger intensity than the $\boldsymbol{Q}_C$ modulation, are observed at domain walls. Their wavevectors ($\boldsymbol{Q}_{DW}$) are half of the sum of two adjacent $\boldsymbol{Q}_C$ (see insets). Based on the pre-determined relation between $\boldsymbol{Q}_{SS}$/$\boldsymbol{Q}_C$ modulations [Figs. 3(g) and (j,k)], we can derive the spin structures of these domain walls, illustrated by superimposed arrows in Figs. 4(c,d). Clearly, each meron/antimeron is associated with a local charge extremum. To quantify it, we calculate the bound charge distribution ($\rho_b = -\nabla \cdot \boldsymbol{P}$) based on the spin textures and the spin-current theory [8], the results are shown in Figs. 4(e,f). A good agreement to experimental maps is observed. Here the 120° wall is formed by two spin spirals with the same chirality; while the 60° wall is formed by opposite chiralities. The corresponding meron/antimeron pairs generate bound charges with opposite signs [marked by rectangles in Figs. 4(e,f)].

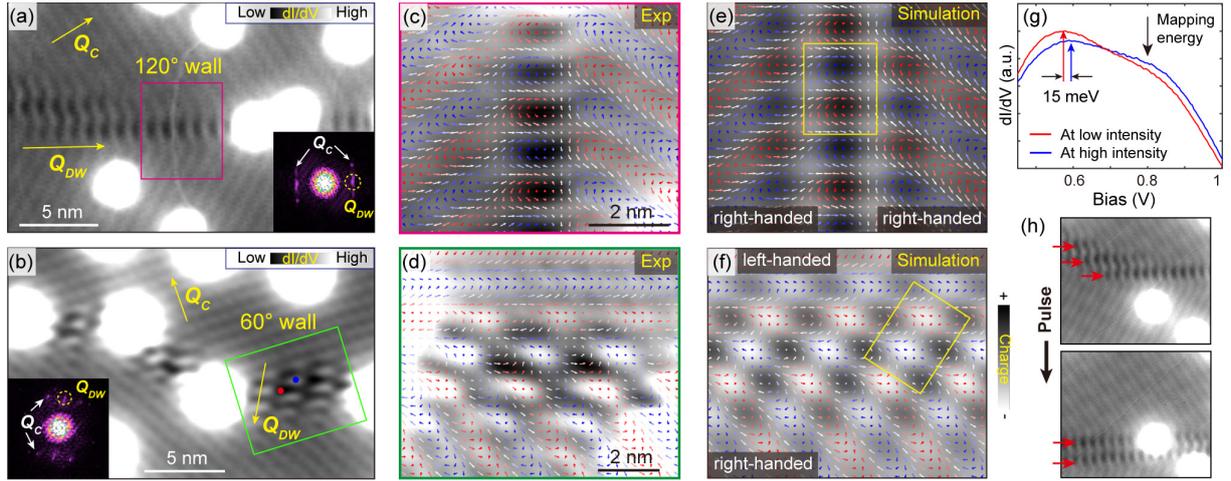

**FIG. 4.** (a, b) dI/dV maps taken around 120°/60° domain walls, respectively ($V_b$ = 0.8 V). Insets: FFT images. (c, d) Zoomed-in images of the marked regions in (a, b), with simulated spin textures superimposed. (e, f) Simulated bound charge distribution of (c, d), based on spin-current theory. The marked regions contain a meron/antimeron. (g) Typical dI/dV spectra taken at the high/low intensity position of $Q_{DW}$ modulation [red/blue spots in (b)]. (h) Two dI/dV maps taken at the same region before and after tip pulse of 4.0 V.

Fig.4(g) shows typical dI/dV spectra measured at high/low intensity positions of $Q_{DW}$ modulation [indicated in Fig. 4(b)]. The spectrum of high-intensity position (blue curve) shows a band shift towards higher energy, suggesting a negative charge accumulation with respect to low-intensity positions (red curve). The relative band shift is ~15 mV, significantly larger than that measured on $Q_C$ modulation [Fig. 2(g)]. As indicated by more accurate MC simulation below [Fig. 5(d)], the net bound charge induced at the domain walls is non-zero. Therefore, electrically charged domain walls associated with topological spin textures are formed. We further found that a moderate tip pulse ($V_b$ ~ 4.0 V) can induce the motion of domain wall [Fig. 4(h)]. This demonstrates possible electrical field control of topological spin textures.

The close relation between observed spin and charge features evidences microscopic magnetoelectric coupling in monolayer $NiI_2$. To gain more insights into the underlying mechanism, we carried out theoretical modeling and simulations. We begin by investigating spin ground state of monolayer $NiI_2$, using a realistic spin model [32] and MC simulations (see Part-I-3 of ref. 41). The Hamiltonian reads:

$$H = \sum_{\langle i,j \rangle_1} \left( J_1 S_i \cdot S_j + K S_i^\gamma S_j^\gamma + B(S_i \cdot S_j)^2 \right) + \sum_{\langle i,j \rangle_3} J_3 S_i \cdot S_j$$

It incorporates the first-nearest-neighbor (1NN) FM exchange $J_1$, third-nearest-neighbor (3NN) AFM exchange $J_3$ [illustrated in Fig. 2(a)], biquadratic interaction $B$, and Kitaev interaction $K$ ($\gamma$ is the Kitaev basis). The above parameters are extracted from previous literature [32]. An enhanced $J_3$ value was used to account for the deviation of spin-spiral propagation in monolayer $NiI_2$ with respected to bulk $NiI_2$.

MC simulations based on this model yield a spin-spiral propagating ~6° away from [110] direction [Fig. 5(a)], consistent with the observed 6.3° deviation. This angle results from competition between [110] and [1$\bar{1}$0] propagating directions, a consequence of Kitaev interaction. The optimized spiral state exhibits a period of 1.85 nm, close to the measured value of 1.74 nm. The simulated spin rotation plane with $\theta = 63.3°$ and $\varphi = 67°$ for both left- and right-handed chiralities also agree with the experimental value ($\theta = 64 - 67°, \varphi = 63 - 65°$). If the Kitaev term is disabled, the spin rotation plane loses its specific anisotropy. This underscores the crucial role of Kitaev interaction in stabilizing the canted spin-spiral [32,51].

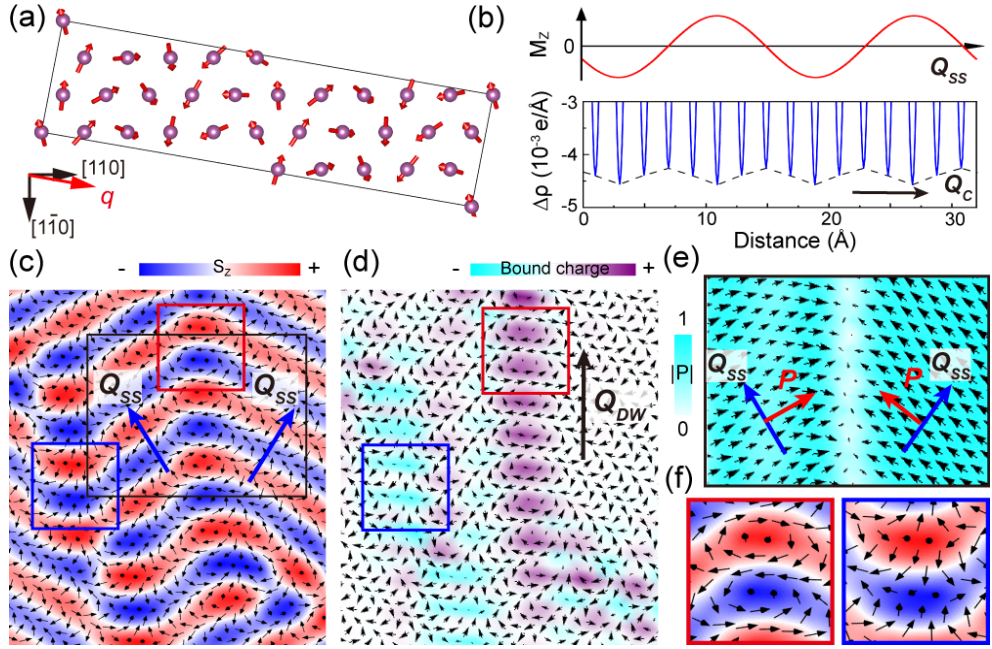

**FIG. 5.** (a) Calculated spin-spiral state of NiI$_2$. (b) The calculated Sz-component of spin-spiral, and the planar averaged differential charge density between spin-spiral and a referenced FM state. (c) Large-scale MC-simulated spin-spiral domains and (d) corresponding GSC model calculated bound charge distribution. (e) Electric polarization (***P***) distribution in the region of black rectangle in (c). (f) Zoomed-in images of the red/blue squares in (c), showing two types of meron/antimeron.

To explore the origin of $\boldsymbol{Q}_C$ modulation, we performed DFT calculations on the charge density for a canted spin-spiral along [110]. Fig. 5(b) (lower panel) shows the charge density difference between this spin-spiral state and a reference ferromagnetic (FM) state. Notably, the differential charge density displays a modulation period exactly half of the spin spiral, and its phase relation with S$_Z$ (upper panel) also agrees well with experimental results in Fig. 3(g). Intuitively, a modulated charge density would generate "modulated electric polarization" in an insulating system such as NiI$_2$, which is consistent with the observed band shift in Fig. 2(g). Meanwhile, because the charge density is obtained by integrating the LDOS over all the occupied states, the LDOS at specific energies may already exhibit a 2$\boldsymbol{Q}$ modulation. Thus, the spin-orientation dependent LDOS modulation previously reported in helical magnetic metal films [43,44], could also contribute to the $Q_C$ modulation here. Furthermore, our calculations demonstrate this $Q_C$ modulation remains robust under changes to spin-spiral period or rotation plane (Fig. S14 of SM [41]), highlighting its universal existence in helical magnetic.

However, the $Q_{DW}$ modulations at domain wall are unlikely to have the same origin with $Q_C$, as they exhibit much stronger intensities and distinct patterns. To investigate the domain wall spin/charge structures more precisely, we performed large-scale MC simulation on 40×40×1 supercells. As shown in Fig. 5(c), multiple spin-spiral domains with $Q$ nearly along equivalent [110] directions are identified. The 120° domain walls are predominant possibly due to relatively low energy. Notably, two types of topological meron/antimeron structures indeed emerge at domain walls, as marked by red/bule squares in Fig. 5(c) and enlarged in Figs. 5(f). We then employed a generalized spin-current (GSC) model [51] to evaluate the electric polarization and bound charge induced by these spin structures. As shown in Fig. 5(e), a nearly uniform ferroelectric polarization ($P$) is induced within single domain, while the polarization undergoes pronounced change at domain boundaries—where topological spin textures present (see Fig. S15 for details [41]). The variations of polarization generate bound charges which predominantly localized at merons/antimerons, as indicated correspondingly in Figs. 5(c) and 5(d). These bound charges give a $Q_{DW}$ modulation similar to that observed in Fig. 4(a). The net bound charge density is calculated to be ~0.05e per nanometers along the domain wall lines.

Now we have presented systematic study on the intricate spin/charge structures, and multiferroicity, in monolayer NiI$_2$. We first determined a canted spin-spiral ground state with full 3D spin resolution. MC simulations incorporating Kitaev interactions highlights its critical role in stabilizing the canted rotation plane. The observation of $Q_C$ charge modulation with a small band shift in dI/dV is consistent with modulated electric polarization or spin-orientation dependent LDOS change [45]. Moreover, meron/antimeron spin textures are directly observed at domain walls, and reproduced by MC simulations. These topological spin textures are accompanied by a stronger charge modulation. Our theoretical models of type-II multiferroicity indicate that they are essentially local bound charges at domain walls. We shall note the observed band shift, ~2.3 mV for $Q_C$ modulation and ~15 mV for $Q_{DW}$ modulation, are significantly smaller than other 2D ferroelectrics like SnTe and Bi monolayer (~100 mV) [52,53]. This is reasonable as the electric polarization in spin-driven multiferroics is typically 1-2 orders weaker than phonon-driven ferroelectrics.

In summary, we obtained the 3D spin structure of monolayer NiI$_2$ by vectorial SP-STM. The fully determined spin-spiral state with canted rotation plane and the accompanied charge modulation—particularly the $Q_{DW}$ modulation at domain wall that associated with topological spin textures, provide microscopic evidence of type-II multiferroicity in such extreme 2D system. Our study also demonstrates the feasibility of electric-field manipulation of topological spin textures, and pave the way for designing low-power spintronic devices containing noncollinear spin orders.

*Acknowledgments*—We thank Prof. Xingao Gong for helpful discussions. This work is supported by National Natural Science Foundation of China (Grants Nos.: 12225403, 92365302, 12188101, 12174060, 12274082), Quantum Science and Technology-National Science and Technology Major Project (Grant Nos.: 2021ZD0302803, 2024ZD0300102), the National Key R&D Program of China (No. 2022YFA1402901), the New Cornerstone Science Foundation, China (Grant No. NCI202211), Shanghai Municipal Science and Technology Major Project (Grant No. 2019SHZDZX01), Shanghai Pilot Program for Basic Research (Grant No. 21TQ1400100), Shanghai Education Committee (Grant No. 24KXZNA01), and the Xiaomi Young Talents Program.

*Data availability*—The data that support the findings of this article are openly available [54].

# Supplementary Materials for
# Microscopic evidence of spin-driven multiferroicity and topological spin textures in monolayer NiI$_2$

Haitao Wang, Tianxing Jiang, Weiyi Pan, Xu Wang, Hongyu Wang, Junchao Tian, Lianchuang Li, Dongming Zhao, Qingle Zhang, Chenxi Wang, Ying Yang, Hongjun Xiang, Changsong Xu*, Donglai Feng*, Tong Zhang*

Corresponding author: csxu@fudan.edu.cn, dlfeng@hfnl.edu.cn, tzhang18@fudan.edu.cn

## Part-I. Experimental Methods:

### I-1: Sample preparation:

The monolayer NiI$_2$ film was grown by molecular beam epitaxy (MBE) method on highly oriented pyrolytic graphite (HOPG). The HOPG substrate was cleaved and followed by outgassing at 400 °C in high vacuum chamber. A single source of NiI$_2$ (99.5% purity powders) was used for the film growth. The source was evaporated from a standard Knudsen cell at 410 °C, while the substrate was kept at 140 °C. The iodine background pressure is $\approx 4 \times 10^{-6}$ mbar and the growth time was 8-10 min. A high iodine background pressure and low growth rate are important for achieving high quality NiI$_2$ film.

### I-2: SP-STM measurements:

Spin-polarized STM experiment was conducted in a cryogenic STM system with a vector magnet (9T-2T-2T) at T = 4.5 K. Spin-resolved tunneling spectra and conductance mapping were measured by Fe-coated tips, which were prepared by depositing 8~10 nm thick Fe film on W tip. Non-spin sensitive measurements were performed with W tips. The W tip was electrochemically etched and flashed up to $\approx$ 2000 K for cleaning before coating. The tunneling conductance (dI/dV) was collected by standard lock-in method and the bias voltage (V$_b$) is applied to the sample.

### I-3: Theoretical modeling and simulation methods:

To describe the magnetic property of monolayer NiI$_2$, we adopt the minimum spin model, as shown below:

$$H = \sum_{\langle i,j \rangle_1} \left( J_1 S_i \cdot S_j + K S_i^\gamma S_j^\gamma + B(S_i \cdot S_j)^2 \right) + \sum_{\langle i,j \rangle_3} J_3 S_i \cdot S_j$$

The parameters used here are extracted from previous literature [1]. The $J_3$ was enhanced from 2.25 meV in [1] to 3 meV in our work, thus the spin spiral propagation direction would change from the [1$\bar{1}$0] directions in bulk to roughly [110] directions detected in the experiment. Based on such spin model, we use Monte-Carlo (MC) simulation to obtain the classical spin textures in monolayer NiI$_2$. Different supercell sizes and shapes are employed to ensure an accurate description of both the period and propagation direction of the canted spin spiral. Moreover, to arrive at large-scale magnetic textures shown in Figs. 5(c, d), we use 40×40×1 supercell. 60000

MC steps are set during our calculations. After MC simulations, a conjugate gradient optimization with the energy convergence criteria being $10^{-6}$ eV is applied to further optimize the spin textures, so that the obtained final spin states are all located at energy minima.

To calculate the $2\boldsymbol{Q}$ charge modulation in spiral states, we perform density functional theory (DFT) calculation using VASP code with the projector augmented wave (PAW) method. During our calculation, we use PBE functional with the inclusion of $U = 4.2$ eV and $J = 0.8$ eV on Ni-3$d$ orbitals, as being adopted in previous work [2]. An energy cutoff of 400 eV is used. For the calculation of unit cell, the $k$-point of 18×18×1 is used. The energy and force convergence criterion are set to be $10^{-6}$ eV and 0.001 eV/Å, respectively.

For the calculation of spin-induced polarization at the domain wall, we use the generalized spin current (GSC) model, which expresses the spin-induced polarization in the form of [3]:

$$\boldsymbol{P} = \sum_{\langle ij \rangle_n}^{n=1,3} M_{ij}^n (\boldsymbol{S}_i \times \boldsymbol{S}_j)$$

Where $M_{ij}^n$ denotes a 3×3 matrix corresponding with the $n$th nearest neighboring (NN) spin pair. Based on a previous study, for the bond along x direction, we adopt $M_{xy}^1 = 109$, $M_{xz}^1 = 139$, $M_{xy}^3 = 164$ (in the unit of $10^{-5}$ eÅ) [4]. With the polarization from GSC model, we can evaluate the bound charge with: $\rho_b = -\nabla \cdot \boldsymbol{P}$, where $\boldsymbol{P}$ denotes the real-space distribution of spin-induced polarization.

## Part-II. Additional STM data and data processing details

### II-1: Surface defects of monolayer NiI$_2$.

Fig. S1 shows additional STM images of 1ML NiI$_2$ on HOPG. At negative $V_b$ of -0.9 V [Fig. S1(b)], point defects can be visualized at the positions of "dark spot" imaged at positive $V_b$ [Fig. S1(a)]. These defects are triangular shaped and centered at the Ni cites, likely to be Ni substitutional defects. We found that the defects do not obviously affect the nearby spin spiral state.

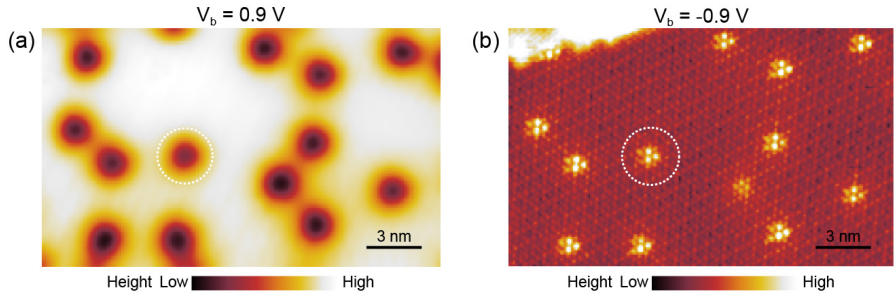

**FIG. S1.** (**a, b**) Topographic images of 1ML NiI$_2$ on HOPG, acquired at $V_b$ = 0.9 V and $V_b$ = -0.9 V, respectively. Point defects can be resolved at negative $V_b$, which are centered at Ni site.

## II-2: Determining the wavevector of spin spiral.

The wavevector of spin spiral ($Q_{SS}$) is determined by performing Fourier transform (FFT) and extracting $Q$ components in reciprocal space. This process is illustrated in Fig. S2 below. Fig. S2(a) is a topographic image showing spin spiral modulation and Fig. S2(b) is the corresponding FFT image. The small deviation of $Q_{SS}$ to the [110] direction (6.3°) is observable in both real-space and FFT images. The in-plane components of $Q_{SS}$ are measured to be 0.13 $a^*$ and 0.09 $b^*$, as illustrated in Fig. S2(b) inset ($a^*$ and $b^*$ are reciprocal basis vectors). The out-of-plane $Q$ component is taken as zero, giving $Q_{SS}$ = (0.13, 0.09, 0).

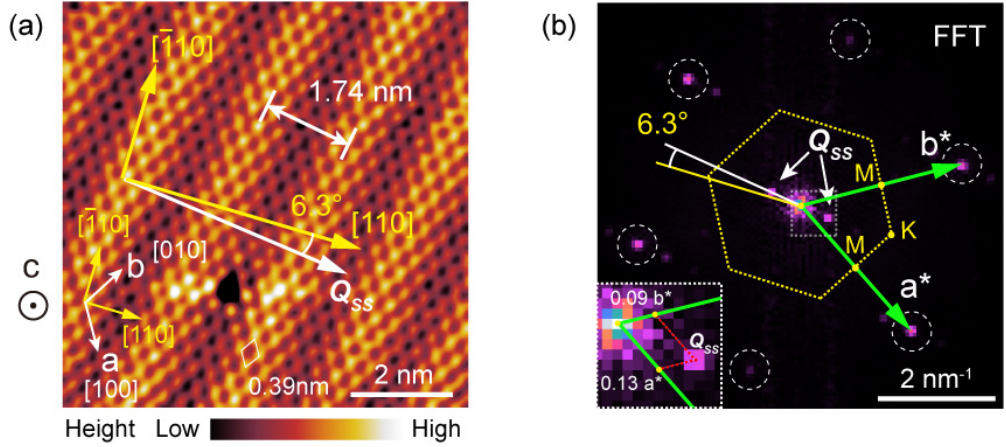

**FIG. S2. Determining the wavevector of spin spiral.** (**a**) Topographic image with spin spiral modulation. The lattice basis vectors (**a** and **b**) and lattice directions are indicated. (**b**) The FFT of panel (a). The reciprocal lattice vectors **a\***, **b\*** and the wavevector of spin-spiral ($Q_{SS}$) are indicated. Dashed hexagon is the first Brillouin zone (BZ). Inset: zoomed-in image showing projected components of $Q_{SS}$ on **a\*** and **b\***

## II-3: Determining the 3D spin structure of spin spiral.

Using an Fe-coated tip magnetized by vectorial magnetic field, we can obtain pure spin signal along any specific direction. By measuring the spin components along X, Y, Z directions, the three-dimensional (3D) spin structure of $NiI_2$ can be determined, including the spin-rotation plane and chirality of spin spirals.

As shown in Fig. S3(j), we choose a region with relatively low defect density to measure the spin structure (marked by dashed rectangle). The position of this region in different dI/dV maps can be precisely determined by aligning its nearby defects. To obtain spin contrast along X direction, dI/dV maps under in-plane fields of $B_X = \pm 0.5$ T were measured, as shown in Figs. S3(a, b). The stripes in these maps show reversed phase, indicating they are from spin signal. Then pure spin contrast along X can be calculated by:

$$S_X = \frac{dI/dV_{(B_x)} - dI/dV_{(B_{-x})}}{dI/dV_{(B_x)} + dI/dV_{(B_{-x})}},$$

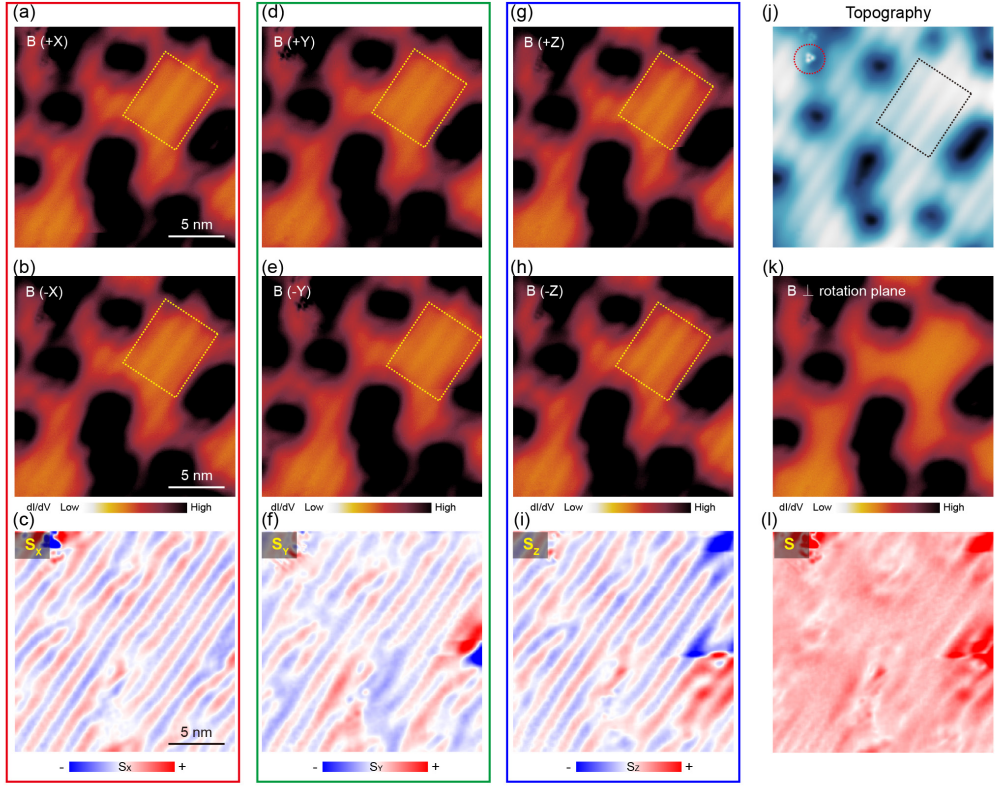

**FIG. S3.** (**a,b, d,e, g,h**) dI/dV maps taken in the same region by Fe-coated tip under $B_X = \pm 0.5$ T, $B_Y = \pm 0.5$ T, $B_Z = \pm 0.5$ T, respectively ($V_b = 0.9$ V, $I = 100$ pA). (**j**) Topographic image of the region. (**k**) dI/dV map taken under $B = 1$ T along the normal direction of the spin rotation plane. (**c, f, i**) Spin-contrast maps ($S_X$, $S_Y$, $S_Z$) obtained by the relative difference of the dI/dV maps of $B_X = \pm 0.5$ T, $B_Y = \pm 0.5$ T, $B_Z = \pm 0.5$ T. (**l**) Total spin magnitude ($|S|$) map calculated from the spin components in (c, f, i).

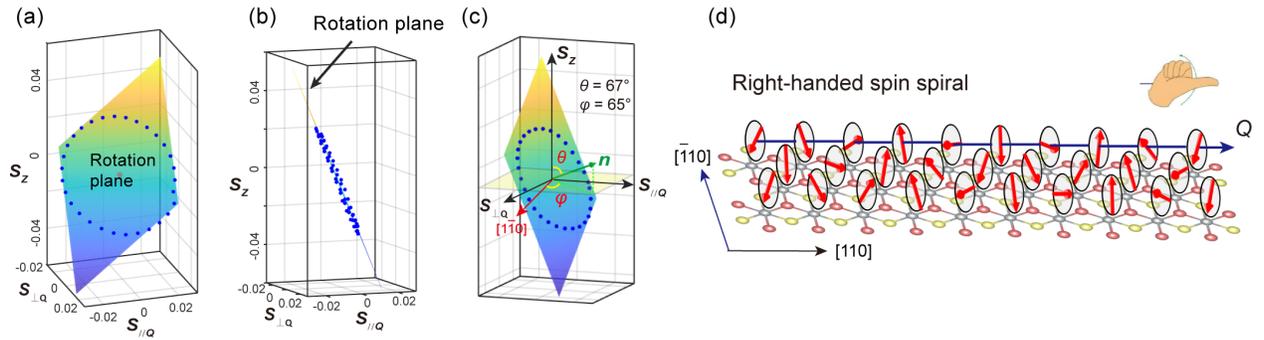

**FIG. S4. Determining the rotation plane and chirality of spin spiral.** (**a, b**) Top and side views of the spin rotation plane in a 3D spin space. Blue dots are the spin vectors extracted from $S_X$, $S_Y$, $S_Z$ maps. They mostly reside in the same plane and formed a circle. (**c**) The polar and azimuthal angles of the normal direction of spin-rotation plane. (**d**) Real-space spin structure of NiI$_2$, a right-handed spin spiral is directly identified.

The resulting $S_X$ map is presented in Fig. S3(c). Similarly, spin contrasts along the Y and Z directions can be obtained, as shown in Figs. S3(d, e, f, g, h, i). Additionally, we have calculated the total spin magnitude: $S = \sqrt{S_X^2 + S_Y^2 + S_Z^2}$, which is nearly uniform, as shown in Fig. S3(l). Therefore, it is reasonable to assume the Fe tip has similar spin sensitivity when it is magnetized along different directions. By combining the spin components in three directions, we can obtain the amplitude and direction of the spin vectors at each position.

To determine the spin rotation plane, we plot all the spin vectors (averaged along the wavefront direction ($\perp Q$) of spin spiral) in a 3D spin space, as shown in Fig. S4(a,b) for different view angles. It's seen that the spin vectors mainly distributed on a circle, indicating they are rotating in the same plane with a nearly constant amplitude. A standard plane fit to this circle yields the polar/azimuthal angles of the normal direction, as indicated in Fig. S4(c). The fitting error is within the range of ±1°. To verify the accuracy of obtained spin rotation plane, we magnetized the tip along the normal direction of the rotation plane. As expected, the spin contrast is barely seen in this case [Fig. S3(k)].

The chirality of the spin spiral can be resolved by plotting the spin vectors in real-space. Fig. S4(d) shows the 3D spin structure of the NiI$_2$ domain in Fig. S3, a right-handed spin spiral is directly identified. We also observed the left-handed spin spiral domain in the sample, as shown in Fig. S5. The 3D spin structure of this domain is also determined by vectorial SP-STM measurement [Fig. S5(a-d)]. It displays a spin-rotation plane with $\theta = 64°, \varphi = 63°$ [Fig. S5(e)], close to that of the right-handed domain [Fig. S4(c)]. The coexistence of right-/left-handed spin spiral indicates an energy degeneracy for different chiralities. This is consistent with our spin model of NiI$_2$ (see Part I-3) and most theoretical works, which suggest the spin spiral in NiI$_2$ originated from the competition between FM ($J_1$) and AFM ($J_2/J_3$) exchange interactions, rather than the DM interaction which is chirality selective. The coexistent of right-/left- handed domain is also necessary to produce different types of meron/antimerion at domain walls and give rise to bound charges with different sign.

We also checked that the spin spiral itself was not affected by the magnetic field used for tip magnetization. In Fig. S6(a, b) we show the phase and line profiles of spin spiral modulation measured under the field of $B_Z$ = -1.5 T to $B_Z$ = +1.5 T. The modulation undergoes a phase reverse at $B_Z \sim 0$ T, which is caused by the flip of tip magnetization. Except the tip spin flip, the spin spiral maintains the same phase up to $|B_Z|$ = 1 T, indicating its stability under this field range.

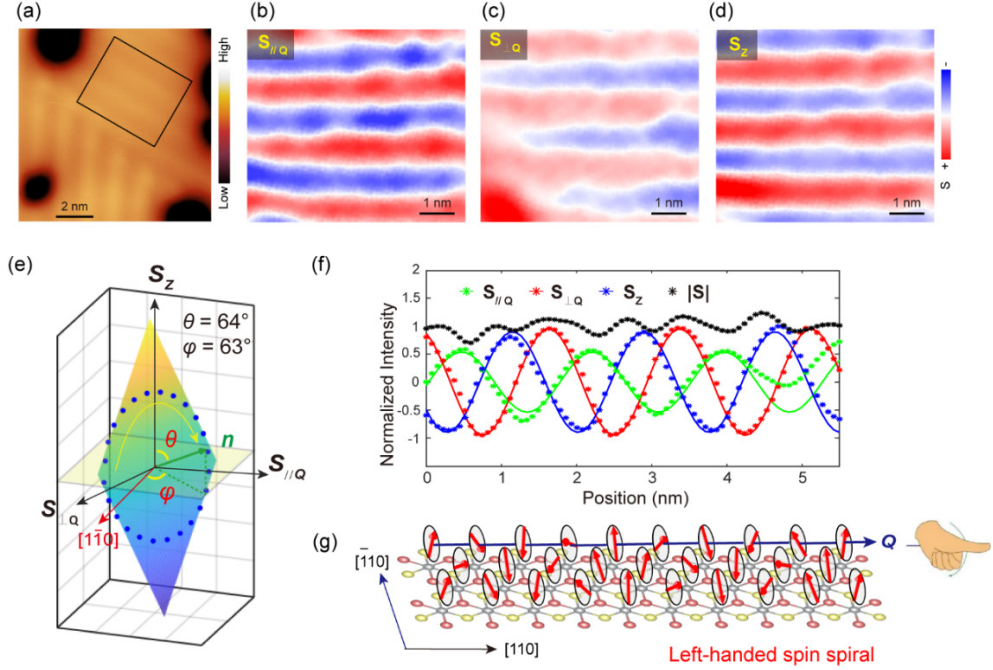

**FIG. S5. The left-handed spin spiral domain.** (a) Topographic image of a monolayer NiI$_2$ surface, (b, c, d) Sx, Sy, Sz maps taken at the marked region in panel (a). (e) Spin vectors obtained from Sx, Sy, Sz maps, which formed a circle in 3D spin space (blue dots). A plane fit yields polar/azimuthal angles of the normal direction of spin-rotation plane (64° and 63°, respectively). (f) Line profiles of the three spin components and the total spin magnitude. (g) Real-space spin structure mapped on the NiI$_2$ lattice, which displays a left-handed spin spiral.

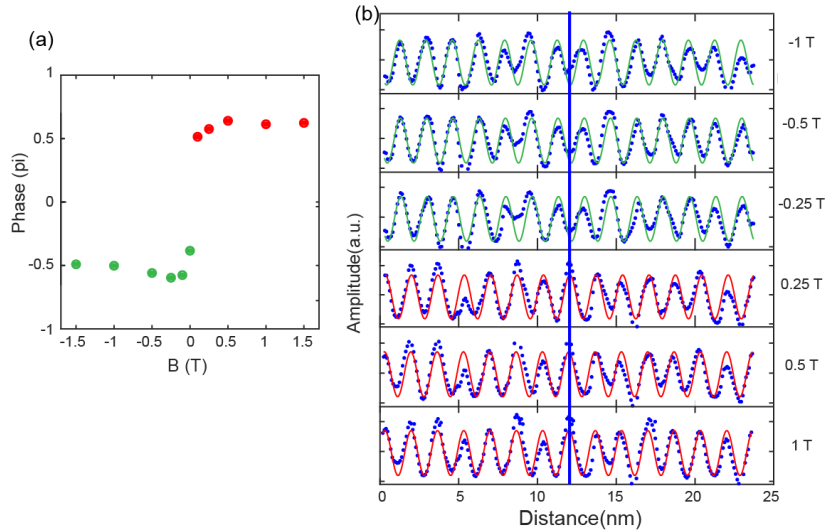

**FIG. S6.** (a) B$_Z$ dependence of the measured phase of spin spiral (extracted from panel b). A $\pi$ phase shift happens when B$_Z$ crossed 0 T. (b) Line profiles of the spin spiral modulations measured under B$_Z$ = -1 T to B$_Z$ = +1 T. The modulation undergoes a phase reverse when B$_Z$ cross 0 T, as tracked by the bule line.

## II-4: Visualizing the 3D spin structure at domain wall (meron/antimeron pair)

The 3D spin structure of spin-spiral domain wall was measured and the results are summarized in Fig. S7. Fig. S7(a-c) show the Sy, Sx, and Sz component maps of a 120° domain wall, which were obtained from raw spin maps [Fig. S7(k)] taken under By = ± 0.5 T, Bx = ± 0.5 T, and Bz = ± 0.5 T, respectively. Fig. S7(d-e) show the MC simulated Sy, Sx, and Sz maps, which agree with the experimental data. Fig. S7(g) shows the obtained full spin structure of this domain wall. Topological meron/antimeron can be directly recognized, which are similar to the MC simulation [Fig. S7(h)] and the typical meron/antimeron illustrated in Fig. S7(j).

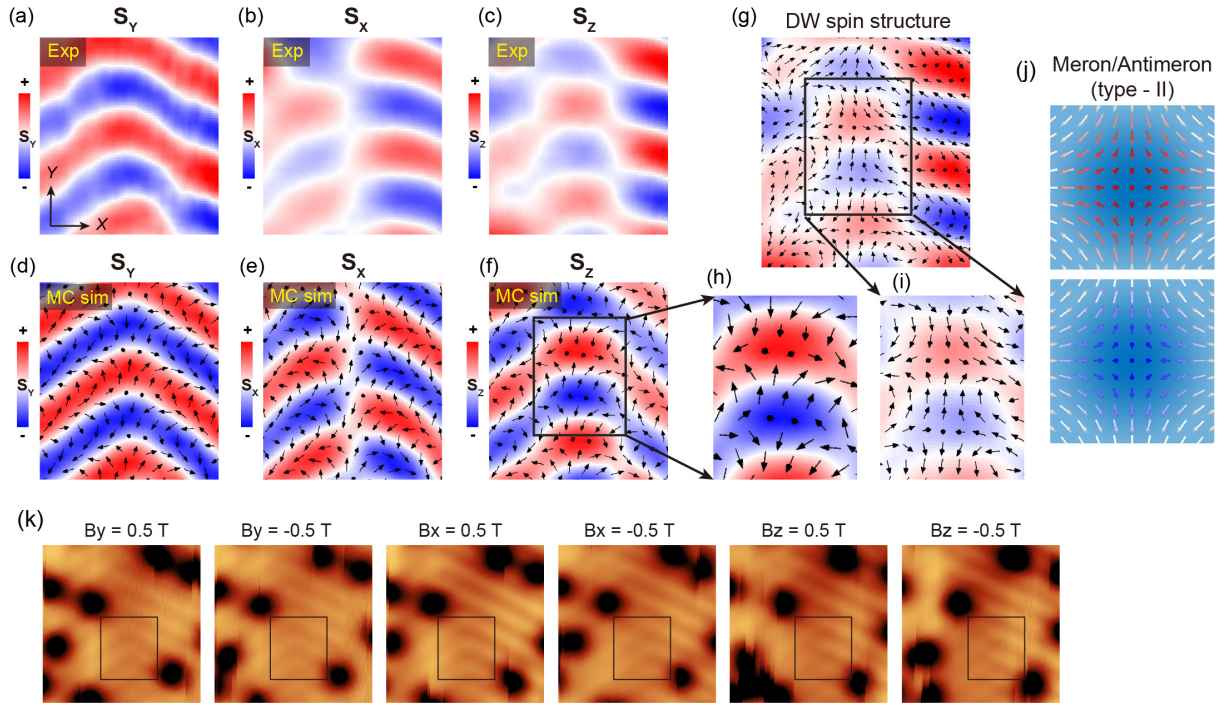

**FIG. S7. Visualizing topological spin textures at the domain wall.** (**a, b, c**) Sy, Sx, and Sz maps taken around a 120° domain wall. (**d, e, f**) MC simulated Sy, Sx and Sz maps of 120° domain wall. (**g**) The full spin structure of 120° domain wall (arrows represent in-plane spin components). (**h, i**) Zoomed-in images of the marked regions in panel (f) and (g), showing the meron/antimeron structures. (**j**) The typical structure of second-type meron/antimeron. (**k**) Raw spin-polarized images measured under By = ± 0.5 T, Bx = ± 0.5 T, Bz = ± 0.5 T. The Sy, Sx, and Sz maps shown in (a-c) are extracted from the marked region.

## II-5: The relation between spin spiral and 2*Q* charge modulation.

Using a spin-polarized tip, the spin spiral modulation and 2*Q* charge modulation can be resolved simultaneously at certain bias voltage, indicative of their direct connection. In Figs. S8(a, b) we show two dI/dV maps taken at $V_b$ = 0.8 V, under $B_Z$ = ±1 T, respectively. Spin modulation can be seen in the difference of the two maps [Fig. S8(c)] while the charge modulation can be seen in their summation [Fig. S8(d)]. Figs. S8(e, f) display the phase relation of these two modulations.

Figs. S9(a, b) show the detailed dI/dV spectra taken along a line across 2**Q** charge modulation. The relative shift of the conduction band peak between high/low intensity positions is more clearly shown in Fig. S9(c). Fig. S9(d) shows the relative shift as function of distance, where an oscillation is clearly observed. These observations are consistent with a modulated band banding effect induced by modulated electric polarization.

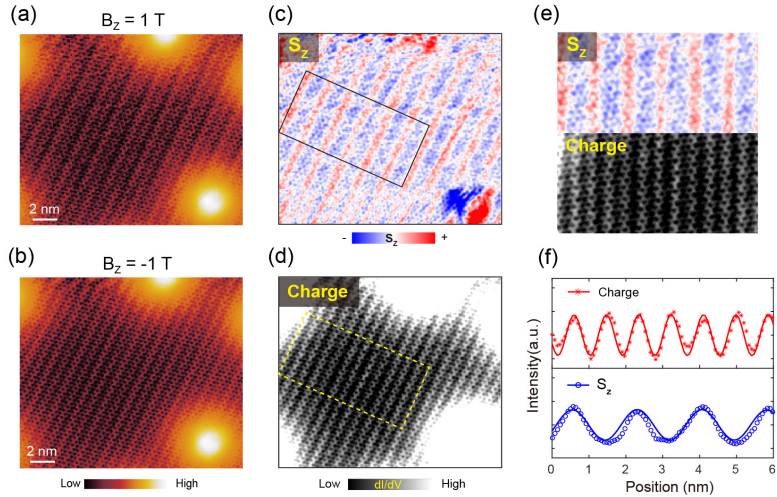

**FIG. S8.** (**a, b**) dI/dV maps taken with a Fe-coated tip, under $B_Z = 1$ T and $B_Z = -1$ T, respectively. (setpoint: $V_b = 0.8$ V, I = 100 pA). (**c, d**) the sum and difference of (a, b), respectively. (**e, f**) the phase relation between $S_Z$ and charge modulation.

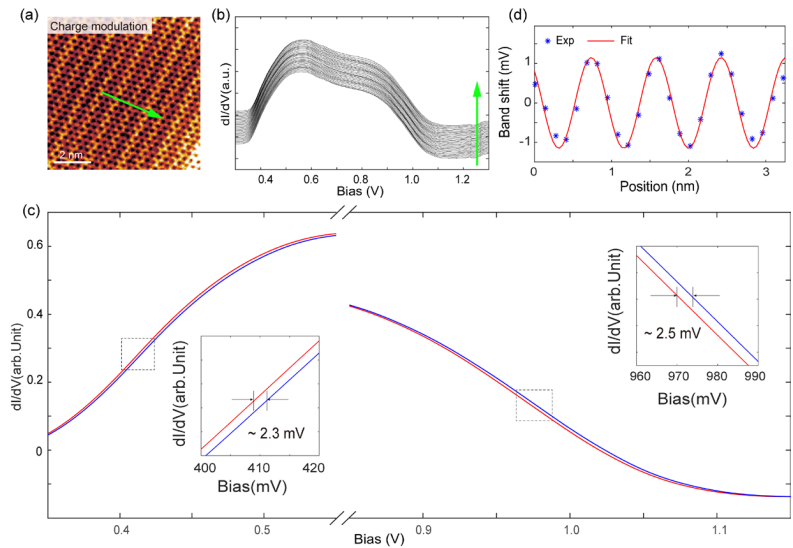

**FIG. S9.** (**a**) dI/dV map showing 2**Q** charge modulation. (**b**) A series of dI/dV spectra taken along the arrow in (a), focusing on the energy range of conduction band peak. (**c**) Enlarged plot of the dI/dV spectra measured at the high (blue curve) and low (red curve) intensities of 2**Q** charge modulation. The insets clearly show the shift of conduction band peak. (**d**) The relative band shift as function of distance.

**Part-III. Domain wall spin/charge structure simulated by the superposition of two spin-spirals.**

To have an intuitive understanding on the domain wall spin structure, we first carried out a straightforward simulation by adding two spin spirals with a finite decay length at domain wall (which generates a double-$Q$ spin spiral state). The validity of this method was verified by more accurate MC simulation shown in **Part-IV**. However, compared to MC simulation, this method can easily simulate large scale domain walls with pre-defined orientation and width. We note that several studies on the helical magnets and row-wise antiferromagnet [5-8] also treated the domain wall structure as a superposition of multiple single-$Q$ state since the energy of multiple-$Q$ state is close to single-$Q$.

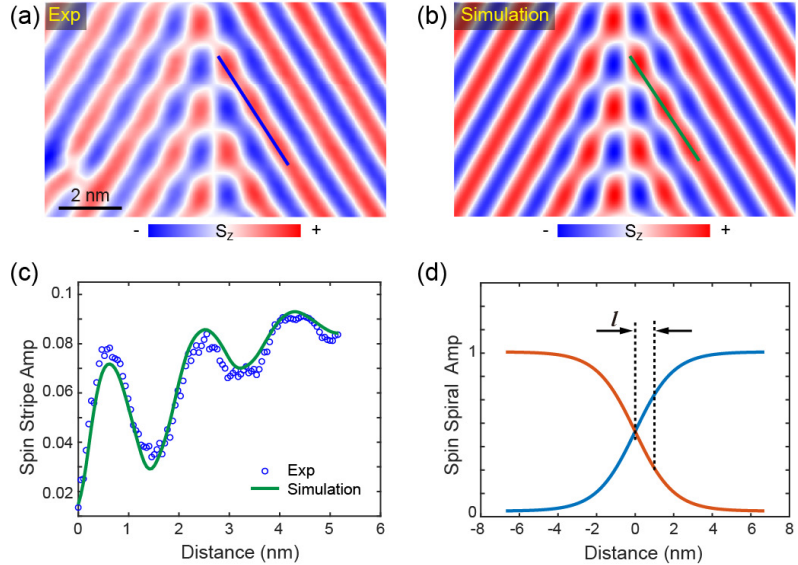

**FIG. S10. (a, b)** The S$_Z$ spin contrast maps obtained from experiment and simulation, respectively. A decay length of $l_0 = 1.0\ nm$ is used for the simulation. **(c)** Line profiles taken along the lines in panel (a, b). **(d)** The decay function $A(x)$ at the domain wall of the two spin spirals used for simulation.

We first simulate the S$_Z$ component at the domain wall. In a spherical coordinate, a (right-handed) spin spiral with a canted rotation plane and a wavevector $Q$ lies in XY plane, is expressed as:

$$\boldsymbol{M} = AM\{\boldsymbol{e_1}[-\sin(\boldsymbol{q}\cdot\boldsymbol{r})\sin\theta\cos\varphi - \cos(\boldsymbol{q}\cdot\boldsymbol{r})\sin\varphi] \\ + \boldsymbol{e_2}[-\sin(\boldsymbol{q}\cdot\boldsymbol{r})\sin\theta\sin\varphi - \cos(\boldsymbol{q}\cdot\boldsymbol{r})\cos\varphi] + \boldsymbol{e_3}\sin(\boldsymbol{q}\cdot\boldsymbol{r})\cos\theta\}$$

Here $M$ is the absolute value of the magnetic moment, $A$ is the normalized amplitude of spin-spiral. $\boldsymbol{e_1}, \boldsymbol{e_2}, \boldsymbol{e_3}$ are the unit vectors pointed along $\boldsymbol{Q}$, perpendicular to $\boldsymbol{Q}$ (in-plane) and along Z direction, respectively. $\theta, \varphi$ are the polar and azimuthal angles of spin rotation plane [see Fig. 3(e)]. To describe the decay behavior of a single-$Q$ spin spiral at domain wall, we used a decay function $A(x) = 1/[1 + e^{(x-x_d)/l_0}]$, where $x_d$ is the center position of the domain wall, $l_0$ is the characteristic decay length. Fig. S10(a) shows a measured $S_Z$ distribution at a 60° domain wall. We can observe decayed modulation on both sides of the domain wall. Using $\theta = 67°$ and $\varphi = 65°$ determined in Part II-3, and $\boldsymbol{Q} = (0.13, 0.09, 0)$, we can simulate the domain wall $S_Z$ component by adding two spin spirals multiplied by the decay function $A(x)$, as shown in Fig. S10(b). By fitting the line profile along the simulated spin stripes to the experimental data [Fig. S10(c)], we can obtain $l_0 \sim 1.0$ nm. Fig. S10(d) shows the corresponding decay function of the two spin-spiral at domain wall.

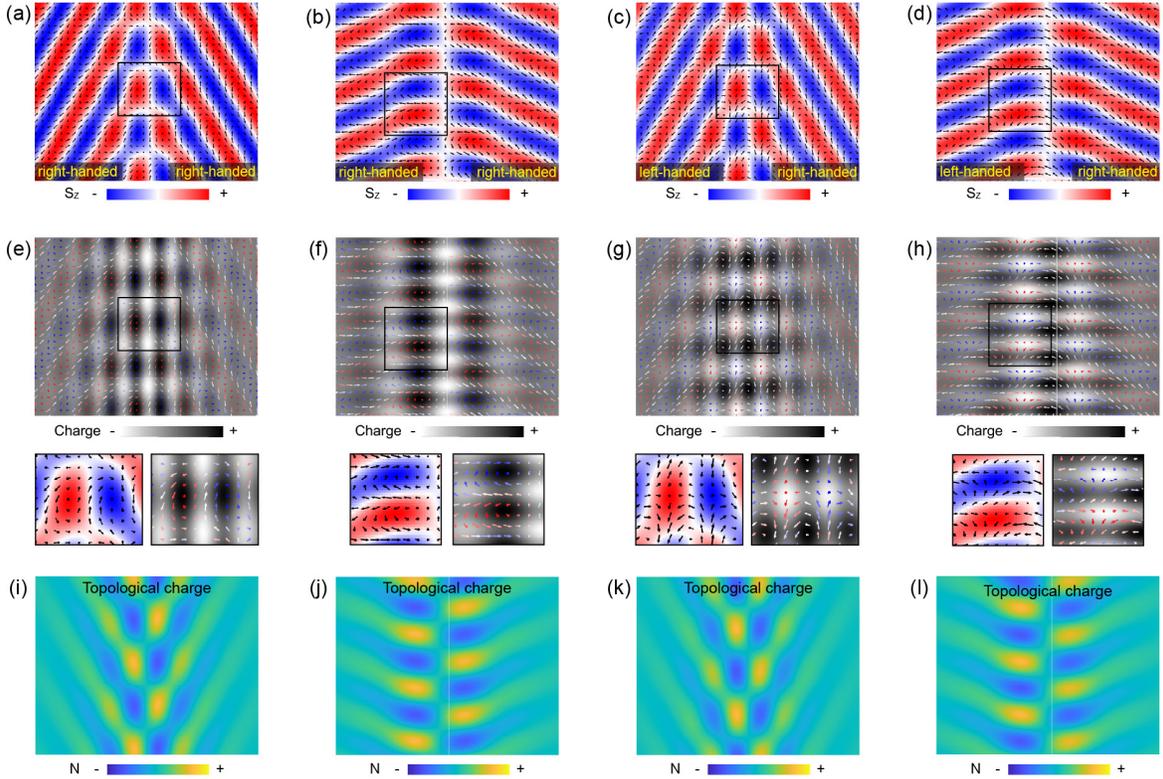

**FIG. S11**. (**a-h**) The spin and bound charge distribution of 60° and 120° domain walls, simulated by superposing two spin spirals with the same and opposite chirality. The rectangular regions show meron/antimeron, as more clearly seen in the enlarged image below. (**i-l**) The topological charge distribution calculated for (a-d), respectively.

We then consider the full spin structure at the domain wall, generated by two spin spirals with the same or different chirality (both right- and left- handed spin spirals have been observed). As shown in Figs. S11(a-d), we simulated the 60° and 120° domain walls by superposing spin spirals of the same and opposite chirality, respectively (indicated in the images). It's seen that when the chiralities of the two spin spirals are the same, they produce vortex-like meron/antimeron at the domain wall, as marked by rectangular region in Figs. S11(a-d). When the chiralities of the two spin spirals are different, they produce another type of meron/antimeron [similar to those shown in Fig. 1(d)]. At the positions of all the meron/antimeron, the topological charge density is non-zero and show opposite sign for meron and antimeron, respectively [Figs. S11(i-l)]. Figs. S11(e-h) show the bound change induced at the domain walls, by using the spin-current model. It is seen that different types of meron/antimeron generated bound charge with opposite sign. These straightforward simulations are consistent with the more accurate MC simulation show in Figs. 5(c, d) and Fig. S15.

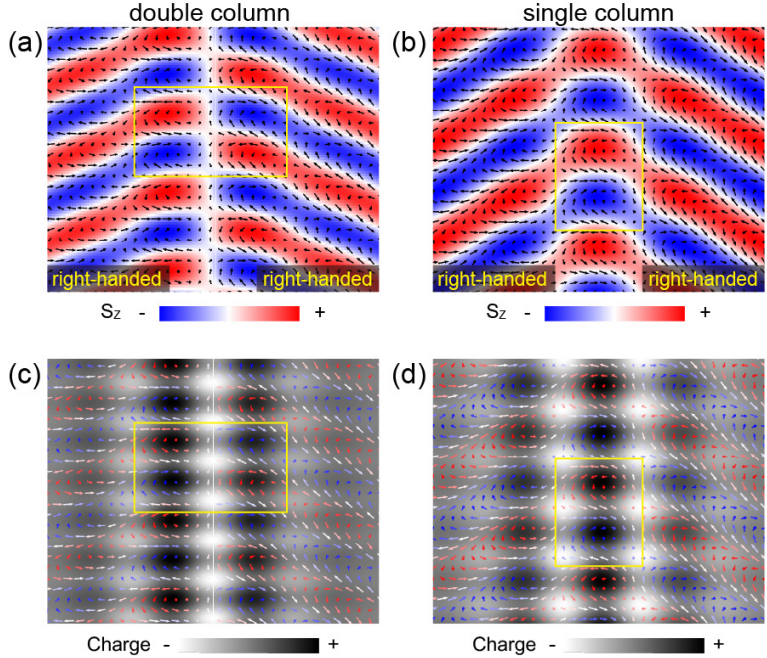

**FIG. S12** (**a, b**) Simulated domain wall structure by two spin spirals with different phase relation at the boundary (dashed lines in the middle). The phase difference is $\pi$ for (a) and 0 for (b). (**c, d**) The bound charge distribution of (a, b), respectively. The boxed regions highlight the meron/antimeron pairs and the associated bound charge.

We further found that the phase difference of the two spin-spirals at the boundary can change the number of meron/antimeron (but does not change their internal structure). As seen in Figs. S12(a, b), a phase difference of $\pi$ generates two columns of meron/antimeron, whereas zero phase difference generates one column. In the experiment, such phase difference is determined by the position of domain boundary. Consequently, when the boundary is shifted by a tip pulse [Fig. 4(h)], the associated charge modulation can change their appearance, as seen in Fig. 4(h).

**Part-IV. Additional results of DFT calculation and MC simulations**

Using first-principles calculations, we computed the electronic density of states (DOS) for monolayer $NiI_2$ with a canted spiral configuration along [110] direction, as depicted in Fig. S13. The results reveal that the valence band is primarily dominated by electrons from I atoms, while the conduction band, characterized by a sharp peak in the DOS, is mainly contributed by electrons from Ni atoms. This DOS profile qualitatively aligns with the experimentally observed DOS of NiI$_2$ in Fig. 2(b) inset. Compared to calculated DOS, the chemical potential in Fig. 2(b) is notably shifted to higher energy, which is possibly due to the interface charge transfer. Moreover, as can be seen in the Fig. S13(c), the magnetization density on Ni atoms, as well as the spin-resolved DOS profile for selected Ni atoms, demonstrate the existence of local spin polarization in spiral state of NiI$_2$, which aligns well with the measurement from SP-STM.

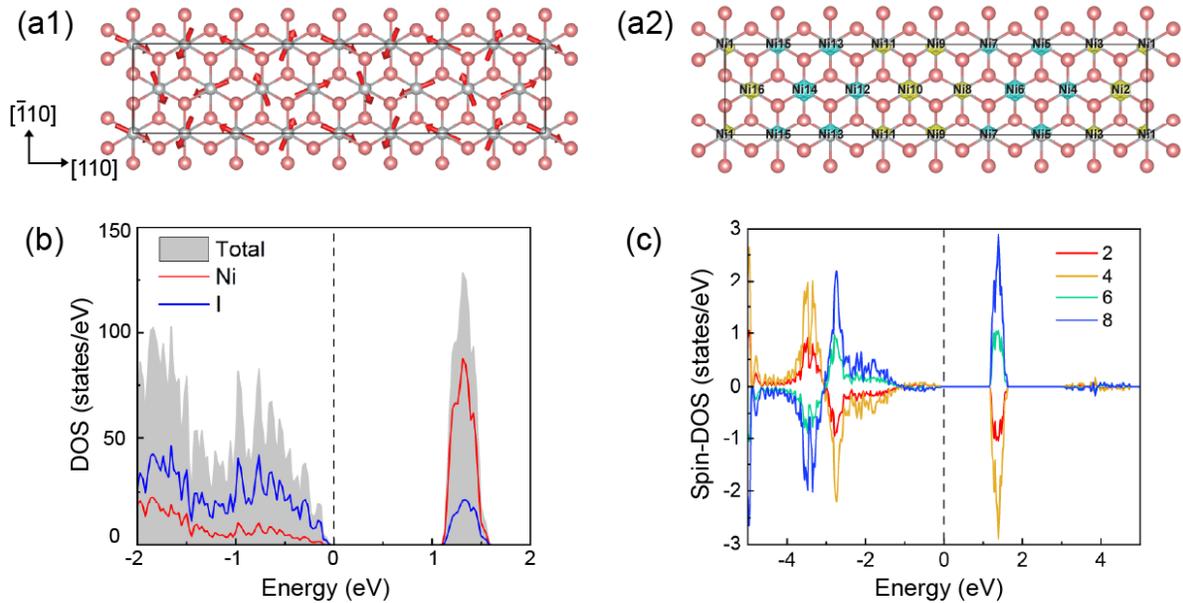

**FIG. S13.** (**a**) The spin spiral in monolayer NiI$_2$ (in a1), as well as the plotted magnetization density along the $z$ component (in a2). (**b**) The calculated DOS, as well as element-resolved DOS in the spiral phase of monolayer NiI$_2$. (**c**) The spin-resolved DOS for $z$ component of spins on selected 4 Ni atoms, which make up of a spiral period. The corresponding Ni atoms are labelled in (a2).

As shown in the main text, our first-principles calculation unveiled a periodic modulation of charge density in the canted spiral phase, with the modulation period of charge being half of the spiral period. Here, we also construct another spiral state with a different period [i.e. 4.5 times of lattice constant, different with the period of 4 times of the lattice constant used to obtain Fig. 5(b) in the main text], as well as a proper screw state with a different spin spiral plane, with the period of 4 times of the lattice constant. Then, we calculate their differential charge density between these spiral states and ferromagnetic state, respectively. As can be seen in Fig. S14, in all these spiral states, the charge density peaks, which mark the Ni atom positions, vary periodically and have a period which is half of the spin spiral period. As a result, the 2***Q*** modulation character of the charge density is robust against the period and the rotation plane of spin spirals.

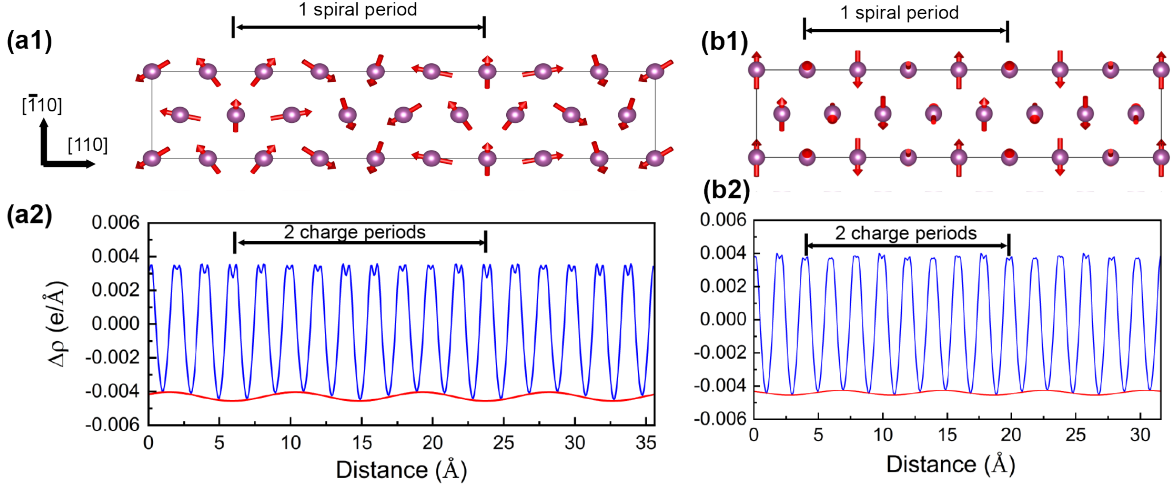

**FIG. S14.** (**a1, a2**) The schematic illustration of a canted spiral state propagating along [110] direction, as well as the calculated differential charge density between this cycloid state and the ferromagnetic state. The period of the spiral state is 4.5 times of the lattice constant, which is 2 times of the charge period. (**b1, b2**) The schematic illustration of a proper screw state propagating along [110] direction, as well as the calculated differential charge density between this proper screw state and the ferromagnetic state. The period of the proper screw state is 4 times of the lattice constant, which is 2 times of the charge period.

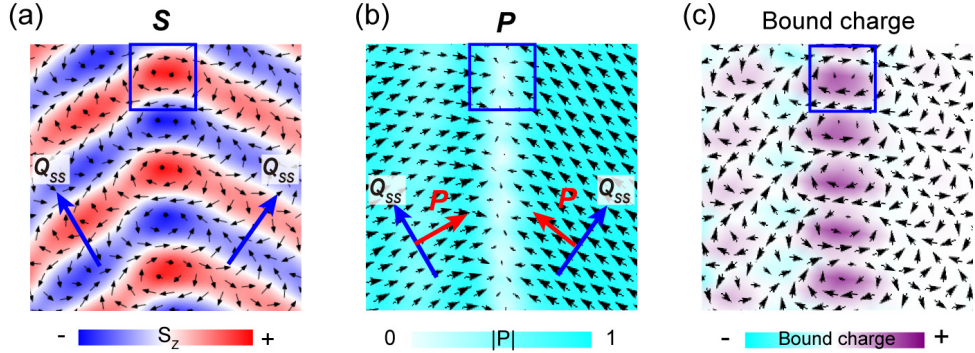

**FIG. S15.** (**a**) MC simulated spin textures of a 120° domain wall and (**b**) corresponding spin-induced electric polarization obtained with GSC model (part I-3). In (a), the black arrows denote in-plane spin component, and the color denotes the Sz component. In (b), the black arrows denote in-plane polarization (***P***), while the magnitude of ***P*** is denoted as blue color. (**c**) The corresponding bound charge distribution of panel (b). Black arrows denote in-plane spin component. Bule squares in panels (a,b,c) mark the position of a meron, a local ***P*** suppression and a bound charge maximum at domain wall, respectively, indicating their close correlation.

Additionally, in Fig. S15 we show more details about spin-induced polarization. Fig. S15(a) is MC-simulated spin textures of a 120° domain wall and Fig. S15(b) is the corresponding electric polarization calculated by GSC model: $\boldsymbol{P} = \sum_{\langle ij \rangle_n}^{n=1,3} M_{ij}^n (\boldsymbol{S}_i \times \boldsymbol{S}_j)$. ***P*** is nearly uniform in each spiral domain, and oriented perpendicular to the spin-spiral wavevector. However, at the domain wall ***P*** is strongly suppressed and exhibits discontinuities, giving rise to localized bound charges via $\rho_b = -\nabla \cdot \boldsymbol{P}$. The resulting distribution of $\rho_b$ in Fig. S15(c) closely reproduces the measured charge map in Fig. 4(a) of the main text. (Note that positive bound charge induces downward band shift and thus low intensity in dI/dV map). It's also seen that the local suppression of $\boldsymbol{P}$ at domain wall actually coincides with the center of topological spin textures and the bound charge maximum, as indicated by the blue squares in Fig. S15(a,b,c) correspondingly. This is because the spins at the center of meron/antimeron are mostly along out-of-plane direction, resulting $\boldsymbol{S_i} \times \boldsymbol{S_j} \to 0$ and thus $\boldsymbol{P} \to 0$ and $\rho_b$ reaches maximum. Therefore, the topological spin texture gives rise to $\boldsymbol{Q}_{DW}$ modulation of $\rho_b$ along domain wall, i.e., each meron/antimeron is accompanied by a local charge maximum.

---